\documentclass[12pt]{article}
\usepackage{epsf}
\newcommand{\eqb}{\begin{equation}}
\newcommand{\eqe}{\end{equation}}
\newcommand{\dmb}{\begin{displaymath}}
\newcommand{\dme}{\end{displaymath}}
\newcommand{\pd}{\partial}
\newcommand{\ep}{\varepsilon}
\newcommand{\eab}{\begin{eqnarray}}
\newcommand{\eae}{\end{eqnarray}}
\newcommand{\ra}{\right\rangle}
\newcommand{\la}{\left\langle}
\newcommand{\e}{\mbox{e}}
\newcommand{\be}{\begin{equation}}
\newcommand{\ee}{\end{equation}}

\newcommand{\La}{\Lambda}
\newcommand{\lsim}{\; ^< \!\!\!\! _\sim \;}

\begin{document}
\begin{titlepage}
\begin{flushright}
MPI-PhT 2001-45 \\
%UMN-TH-1830/99   \\
%%% October 15
\end{flushright}
\vspace{0.6cm}

\begin{center}
\Large{{\bf Gauged Inflation}}

\vspace{1cm}

Ralf Hofmann and Mathias Th. Keil

\end{center}
\vspace{0.3cm}

\begin{center}
{\em Max-Planck-Institut f\"ur Physik\\ 
Werner-Heisenberg-Institut\\ 
F\"ohringer Ring 6, 80805 M\"unchen\\ 
Germany}
\end{center}
\vspace{0.5cm}

\begin{abstract}

We propose a model for cosmic inflation which is based on an effective
description of strongly interacting, nonsupersymmetric matter within
the framework of dynamical Abelian projection and centerization.  The
underlying gauge symmetry is assumed to be $SU(N+1)$ with
$N \gg 1$. Appealing to a thermodynamical treatment, the ground-state
structure of the model is classically determined by a potential for
the inflaton field (dynamical monopole condensate) which allows for
nontrivially BPS saturated and thereby stable solutions. For $T<M_P$
this leads to decoupling of gravity from the inflaton dynamics. The
ground state dynamics implies a heat capacity for the vacuum leading
to inflation for temperatures comparable to the mass scale $M$ of the
potential. The dynamics has an attractor property. In contrast to the
usual slow-roll paradigm we have $m\gg H$ during inflation. As a
consequence, density perturbations generated from the inflaton are
irrelevant for the formation of large-scale structure, and the model
has to be supplemented with an inflaton independent mechanism for the
generation of spatial curvature perturbations. Within a small fraction
of the Hubble time inflation is terminated by a transition of the
theory to its center symmetric phase. The spontaneously broken
$Z_{N+1}$ symmetry stabilizes relic vector bosons in the epochs
following inflation. These heavy relics contribute to the cold dark
matter of the universe and potentially originate the UHECRs beyond the
GZK bound.

\end{abstract} 

\end{titlepage}

\section{Introduction}

Inflationary cosmology was invented to resolve a number of problems
posed by the  conventional hot big bang scenario
\cite{Guth,Linde0}. Experimental facts not explained by the old
cosmology were (A) the apparent spatial flatness of the universe, (B)
the nondetection of magnetic monopoles stemming from the spontaneous
symmetry breaking in grand unified gauge theories (GUTs) at a scale of
$\sim 10^{16}$ GeV, and (C) the horizon problem, arising from the
extreme isotropy of the cosmic microwave background radiation
(CMBR). In addition, it was hoped that the mechanism responsible
for inflation could also provide the small density perturbations
needed to seed large-scale structure formation (D).

Superluminal expansion of the universe by a factor of about $\e^{60}$
at a sub-Planckian epoch can solve the flatness problem (A)
\cite{Linde1}. To meet the constraints on postinflationary density
perturbations, however, the new inflationary cosmology of
\cite{Linde0} needs densities at inflation much smaller than
$10^{-10}\times M^{4}_P$ \cite{LindeB}. This, in turn, would cause the
collapse of a typical closed universe before it could inflate. A
resolution of this problem was proposed by assuming chaotic inflation
\cite{Linde1}. We will argue in this paper that even within the new
inflationary scenario collapse can be avoided if the initial radius
$a_0$ of a closed universe lies above some critical value.

The monopole problem (B) is resolved if inflation sets in at scales
considerably smaller than the GUT scale. In models where inflation is
driven by a single, minimally coupled, classical, real scalar field
(inflaton), either the potential for this field is constructed to
allow for a regime of slow-roll, which implies a small mass $m$ for
the field fluctuations if compared to the Hubble parameter $H$
\cite{Linde0}, or a nonconventional kinetic term for this field is
introduced in the action \cite{Mukhanov}. However, the recent work of
\cite{Linde2} discusses also fast-roll inflation. The realization
within the context of the present work is that slow-roll does not
necessarily imply the hierarchy of mass scales in the inflaton sector
which is expressed by $m\ll H$.

Inflation also solves the horizon problem (C) since it implies that
todays observable universe was entirely contained in a causally
connected patch of space prior to inflation. 

Assuming $m\ll H$ during inflation, the problem of density
perturbations (D) was addressed in the past by considering quantum
fluctuations of the inflaton field which adiabatically freeze into a
scale invariant spectrum of classical fluctuations upon horizon-exit
during inflation. Theses fluctuations, when re-entering the horizon
after inflation, were thought to be responsible for the density
perturbations seeding the formation of large-scale structure. It was
only recently that alternatives to $m\ll H$ have been put forward in
the literature \cite{Lyth}. Post-inflationary density perturbations
are explained by the decay of a moduli field whose existence and
dynamics is independent from the inflaton sector. Our proposed  model
relies on the existence of this extra field but, at the same time,
assures that  moduli fields are not generated as a consequence of  the
inflaton dynamics. This resolves the conventional moduli problem
\cite{Randall,LythStewart}.

We propose a large-$N$, pure $SU(N+1)$ gauge theory to be responsible
for the dynamics leading to inflation and terminating it. The idea is
that at the time when the universe enters the post-Planckian epoch the
physics of matter is dominated by a large gauge symmetry which, in the
course of cosmic evolution, dynamically freezes more and more of its
subgroups to their respective centers hence generating an ever
increasing number of mass scales. The first center transition is the mechanism proposed here to
terminate the first regime of quasi-exponential expansion. We identify
this regime with usual inflationary cosmology. The question how fermions appear and
enter the dynamics will not be addressed in this work.

In order to be able to
access the complicated vacuum structure of $SU(N+1)$ gauge theory we
resort to a conjectured, effective description based on the picture of
a dual superconductor (Abelian Higgs model) at high temperature $T$
\cite{dualsc} and based on confinement of fundamental test charges due
to the condensation of center vortices at low energy \cite{center}
{($Z_{N+1}$ symmetric Higgs sector). On the one hand, this picture is
supported by lattice simulations of $SU(2)$ pure gauge theory
\cite{centerlat}. On the other hand, models based on center-vortex
condensation are quite successful in explaining the low-energy
features of the QCD vacuum \cite{Engelhardt}. In the following 
we will refer to the complex Higgs
field as ``inflaton field.'' 

In the high $T$ regime the surviving gauge group should be maximally
Abelian, i.e. $U(1)^{N}$. This renders the vacuum a condensate of
magnetic monopoles being charged under $N$ different $U(1)$
groups. For simplicity we restrict ourselves to a higgsed model of a
single $U(1)$ group which still is in accord with the qualitative
picture \cite{hof5}. At low energy a spontaneously broken $Z_{N+1}$ 
symmetry will be apparent for the same field $\phi$ that described
monopole condensation at high $T$. Thereby, the distinction of high
and low energy is given by the single mass scale $M$ of the
potential. This potential is constructed to allow for BPS saturated
\cite{BPS} Euclidean ground state dynamics at high $T$. In the course
of cosmological evolution it smoothly enters a regime $T \simeq M$
where a description in Minkowskian signature becomes necessary due to
destruction of thermal equilibrium by transition to the center
symmetry. BPS saturation at high $T$ suppresses the gravitational
coupling term in the second-order (Euclidean) equations for the Higgs
field, topologically stabilizes the dynamics, justifies the classical
(effective) description, and leads to the appearance of a temperature
dependent cosmological constant \cite{hof5}.

Much in contrast to perturbatively obtained effective potentials we 
have no explicit $T$ dependence in our potential. Rather, the $T$
dependence of physical quantities, such as the energy density of the
vacuum, arises entirely from the $T$ dependence of the solutions to
the field equations.

Assuming Einstein gravity to hold, the cosmic evolution of an
isotropic and homogeneous universe is governed by the Friedmann
equations. In our approach the source terms are the black body
radiation of the massive gauge field quanta (scalar quanta are of mass
considerably larger than temperature and hence much heavier than the
vector bosons) and the aforementioned cosmological constant
$\La=\La(T)$. 

The pattern of symmetry breaking induced by the cooling universe is:
Unbroken gauge symmetry at temperatures $T \gg M$ $\rightarrow$
increasing spontaneous gauge symmetry breaking by a growing inflaton
amplitude in an intermediate regime $\rightarrow$ explicit breaking of
the $U(1)$ gauge symmetry to its discrete subgroup $Z_{N+1}$ for field
amplitudes close to $M$. This is a concrete realization of Krauss' and
Wilzcek's observation that fundamental gauge symmetries masquerade as
discrete symmetries for observers at low energies \cite{Wilczek}.

The evolution of the scale factor $a$ can be summarized as: radiation
dominated, slow power-law expansion $\rightarrow$ energy density of
the vacuum becomes comparable with that of radiation implying stronger
power-law expansion $\rightarrow$ quasi-exponential expansion due to
vacuum dominance (inflation) $\rightarrow$ thermal nonequilibrium due to the 
termination of inflation by vacuum decay and creation of new particles
(pre- and reheating). Inflationary relics are vector particles of mass
$\sim M$ whose relative stability is protected by a discrete symmetry
$Z_{N+1}$ \cite{Wilczek,Cohn}.

The paper is organized as follows: In Section~\ref{sec:model} we
introduce the model. Thermal equilibrium prior to and during a large
part of inflation implies a description in Euclidean space with a
compactified time dimension of length $\beta\equiv T^{-1}$. At several
occasions we exploit the fact that observables such as the
energy-density of the vacuum or the Hubble parameter being
time-independent in Euclidean signature have a trivial analytical
continuation to Minkowskian signature. Requiring the inflaton field to
be BPS saturated in a non-trivial way for the regime governed by the
$U(1)$ gauge symmetry, severely constrains the corresponding
potential. Demanding in addition that there is a smooth transition to
the center symmetric regime, which is determined by a single mass
scale $M$, fixes the inflaton potential uniquely. After the
construction of the potential we solve the dynamics of the ground
state which is assumed to be locally Lorentz invariant. As a result,
we obtain a cosmological constant $\La$ depending linearly on
$T$. Four-momentum conservation in a Friedmann universe, whose
evolution is determined by (massive) black body radiation of the gauge
field excitations and $\La=\La(T)$, determines $\La$ as a function of
$a$. We find that the inflaton amplitude at inflation is a fixed point
of cosmic evolution. 

In Section \ref{sec:evolution} we obtain the time evolution of $a$. We
show that all phenomenological requirements for inflationary cosmology
can be met provided an additional light scalar field is invoked to
generate the density perturbations demanded by the CMBR anisotropy and
the constraints imposed by models of large-scale structure
formation. General implementations of this additional scalar have
recently been discussed \cite{Lyth}. Moreover, we prove that a closed
and initially large enough universe of sub-Planckian density does not
collapse but may also inflate. 

An investigation of the physics terminating inflation is performed in
Section~\ref{sec:postinf}. At the point $|\phi|_c$, where the breaking
of continuous gauge symmetry becomes noticeable, i.e. where
$\left.\pd_{|\phi|}^2 V(|\phi|)\right|_{|\phi|=|\phi|_c}=0$ for the first
time, thermal equilibrium is destroyed due to the inflaton amplitude
becoming time  dependent even in the Euclidean and due to tachyonic
excitations. This leads to a termination of inflation by  tachyonic
preheating and the subsequent generation of new matter. Due to a
spontaneously broken $Z_{N+1}$ symmetry there are no Goldstone
modes. Therefore, no isothermal density fluctuations can be produced
from the dynamics driving and terminating inflation. In
Section~\ref{sec:concl} we summarize the results, speculate on
implications of this work, and provide an outlook.

\section{The model}
\label{sec:model}

\subsection{BPS saturated thermodynamics: The ground state}

The central assumption of our model is that very early cosmology effectively is
driven by a gauged $U(1)$ theory of a classical and complex ``inflaton
field.''  The scalar as well as the gauge sector of the model are
taken to be minimally coupled to Einstein gravity. The corresponding
action is
%********
\eqb
\label{act}
S=\int d^4x \sqrt{-g}\left[
- \frac{1}{16\pi G}R - \frac{1}{4} F_{\mu\nu}F^{\mu\nu} +
\overline{{\cal D}_\mu \phi}{\cal D}^\mu \phi -  
V(\bar\phi \phi)\right]\, ,
\eqe
%******** 
where ${\cal D}_\mu\equiv\pd_\mu+ieA_\mu$ denotes the gauge covariant
derivative, and $V$ is an effective potential to be specified
later. Assuming thermal equilibrium, the dynamics is considered for a
compact time dimension in Euclidean signature  $0\le\tau\le\beta\equiv
1/T$.

According to the standard cosmological assumptions of spatial homogeneity and 
isotropy the inflaton field $\phi$ is independent of the spatial
variables but may depend on the Euclidean time coordinate. It will be
shown below that the vacuum is dominated by the BPS saturated dynamics
of $\phi$ alone, which is solved in a gauge allowing 
for physical boundary conditions. Therefore one may regard the gauge field
dynamics in the background of the inflaton solution.

What about gravitational deformation? Gravity is represented by the 
Euclidean version of the Robertson-Walker metric with scale factor
$a$, while $\phi$ and $\bar{\phi}$ satisfy the BPS equations
%********
\eqb 
\label{BPS}
\pd_\tau\phi=\bar{V}^{1/2}\,,\ \ \ \ \pd_\tau\bar{\phi}={V}^{1/2}\ ,
\eqe
%*******
where $V\equiv\bar{V}^{1/2}{V}^{1/2}$. Taking the divergences of the
vector equations
$v_\mu=\delta_{\mu\tau}\pd_\tau\phi=\delta_{\mu\tau}\bar{V}^{1/2}$ and
$\bar{v}_\mu=\delta_{\mu\tau}\pd_\tau\bar{\phi}=\delta_{\mu\tau}{V}^{1/2}$
yields
%*********
\eqb
\label{2or}
\nabla_\mu v_\mu=
\nabla_\tau\bar{V}^{1/2}+3H\bar{V}^{1/2}\ ,\ \ \ \ 
\nabla_\mu \bar{v}_\mu=\nabla_\tau{V}^{1/2}+3H{V}^{1/2}\ ,
\eqe
%*********
where $\nabla_\mu v^\nu=\pd_\mu+\Gamma^\nu_{\mu \kappa}v^\kappa$
denotes the action of the gravitational covariant derivative and
$H\equiv\frac{da/d\tau}{a}$ is the Euclidean Hubble parameter. Since
the covariant time derivatives on the right-hand sides (RHSs) of
eqs.\,(\ref{2or}) act on scalars we can replace them by ordinary time
derivatives. Using the BPS equations (\ref{BPS}) and the definitions
of ${V}^{1/2}$ and $\bar{V}^{1/2}$, we can express the RHSs of
eq.~(\ref{2or}) as
%*********
\eab
\label{RHS} 
\frac{\pd \bar{V}^{1/2}}{\pd \bar{\phi}}{V}^{1/2}+3H\bar{V}^{1/2}&=&
V^{-1/2}\frac{\pd V}{\pd \bar{\phi}} V^{1/2}+3H\bar{V}^{1/2}=
\frac{\pd V}{\pd \bar{\phi}}+3H\bar{V}^{1/2}\,,\nonumber\\ 
\frac{\pd {V}^{1/2}}{\pd {\phi}}\bar{V}^{1/2}+3H{V}^{1/2}&=&
\bar{V}^{-1/2}\frac{\pd V}{\pd \phi} \bar{V}^{1/2}+3H{V}^{1/2}=
\frac{\pd V}{\pd \phi}+3H{V}^{1/2} \ .
\eae
%******** 
Invoking eq.~(\ref{2or}), we realize that $\bar{\phi}$ and $\phi$
would approximately satisfy the Euclidean, second order field
equations if the gravitational coupling terms in eqs.\,(\ref{RHS}) are
small compared to $\pd_\phi V$ for the relevant range 
of $|\phi|\le|\phi|_c$. We show this
in Sec~\ref{sec:decoupling}. Apparently, $\bar{V}^{1/2}$ and
${V}^{1/2}$  are only fixed up to constant phases $\e^{i\delta}$ and
$\e^{-i\delta}$, respectively. This freedom does not survive if the
solution of the ground state dynamics is based on inflaton dominance
(see below).

Let us now construct the potential $V$. We demand $(i)$ the existence
of non-trivial BPS saturated solutions along the compact, Euclidean
time coordinate in the regime of $U(1)$ gauge symmetry and a smooth
transition to the $Z_{N+1}$ symmetric regime. $(ii)$ In addition we
require the potential to be characterized by a single mass scale
$M$. Detailed justifications can be found in \cite{hof5}.

$(i)$ fixes the potential as (see also \cite{Hofmann0})
%********
\eqb
\label{pot}
V(\bar{\phi}\phi)=\frac{M^6}{\bar{\phi}\phi}+\lambda^2 M^{-2(N-2)}
(\bar{\phi}\phi)^N-2\,\lambda M^{5-N}\frac{1}{\bar{\phi}\phi}\mbox{Re}\,\phi^{N+1}\ , 
\eqe 
%********* 
where $\lambda$ denotes a dimensionless coupling constant. In the
limit $N\to\infty$ the potential is gauge invariant for the whole
$\phi$ range. However, the transition to the trivial vacuum is not
smooth, violating $(i)$. For finite $N$ the $N+1$ points of zero
energy are given by the $N+1$ unit roots times
$M/\lambda^{1/(N+1)}$. In order to meet $(ii)$ we have to set
$\lambda=1$. We stress that adding a constant to the potential would
destroy $(i)$. Therefore, the corresponding contribution to the cosmological constant vanishes after the
transition to the center symmetric regime! In section 2.3 we will see
that demanding the inflationary fixed point $|\phi|_i$ to coincide
with the potential's first point of inflexion $|\phi|_c$ (onset of
tachyonic excitations) yields $N\sim 34$. The relative deviation of
the pole part and the full potential at this point is about
1\%. Therefore, even at finite and not too small $N$ and for
$|\phi|<|\phi|_c$ we may regard the potential eq.~(\ref{pot}) as gauge
invariant and given by the pure pole term whereas $Z_{N+1}$ symmetry
is apparent for $|\phi|>|\phi|_c$. A graph of the potential is shown
in Fig.\,1. 
%***************.
\begin{figure}
%\vskip3mm
\vspace{6.2cm}
\includegraphics{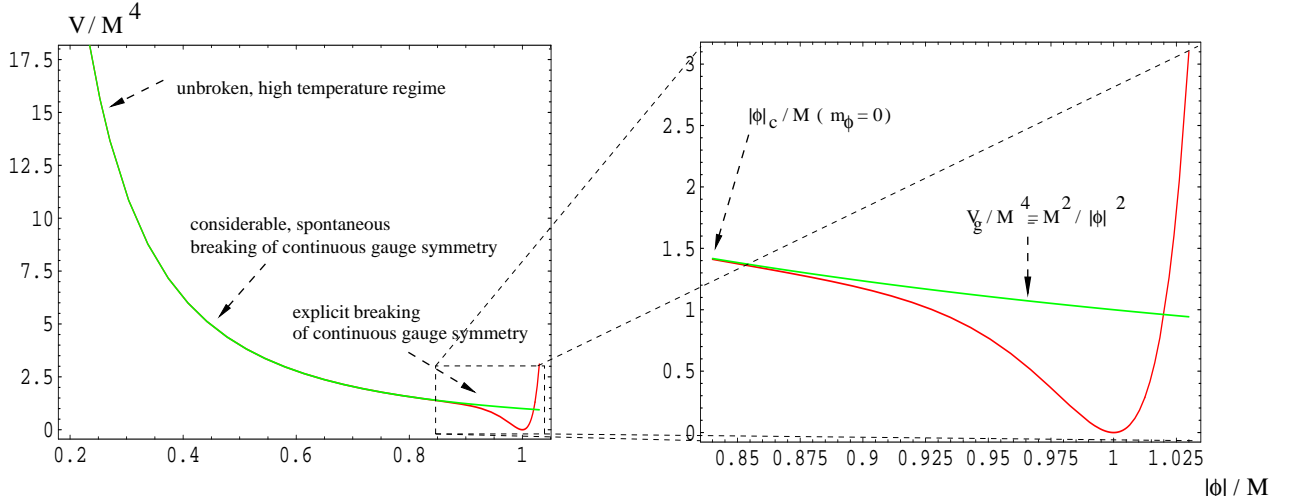}
%\centerline{\epsfbox{numsol.eps}}
\caption{The potential $V(|\phi|^2)$ of eq.\,(\ref{pot}) allowing for
BPS saturated solutions along the Euclidean, compact time
dimension. We have set $N=34$ and $\lambda=1$. In the enlarged region,
where the continuous gauge symmetry is broken explicitly, the gauge
invariant, pure pole part $V_g$ deviates strongly from the full
potential.}
\end{figure}
%*************** 

We now turn to the solution of the ground state dynamics at high
$T$. As mentioned above, our ansatz is a dominant inflaton field. The
idea is to choose the gauge such as to shuffle a maximum of physics
into the scalar sector. In this sense a chosen gauge at finite
temperature is physical if and only if the solution $\phi$ to
eq.~(\ref{BPS}) is periodic.

We now show that the choice of phase in the BPS equations is
correlated with a choice of gauge. For example, imposing unitary gauge
$\phi=|\phi|$ before solving the BPS equations (\ref{BPS}),
$\bar{V}^{1/2}$ and ${V}^{1/2}$ must be real, leading to the following
solution
%********
\eqb
\phi(\tau)=\sqrt{\phi^2_0+2M^3(\tau-\tau_0)}\ .
\eqe
%********
Obviously this is not a periodic function. The only choices
of phase yielding the following periodic solutions \cite{ShifmanDvali}
%**********
\eqb
\label{BPSsol}
\phi^{(n)}(\tau)=\sqrt{\frac{M^3\beta}{2|n|\pi}}\,
\e^{2n\pi i\frac{\tau}{\beta}}\ ,\ \ (n\in{\bf Z})\ 
\eqe
%********* 
correspond to 
%*******
\eqb
\label{V1/2}
V^{1/2}(\phi)=\pm i\frac{M^3}{\phi}\,,\ \ \ \ \bar{V}^{1/2}(\phi)=\mp i\frac{M^3}{\bar{\phi}}\,,\ \ 
\eqe
%*******
for $(|\phi|<|\phi|_c)$.
In eq.~(\ref{BPSsol}) the integer $n$ counts the number of times the
corresponding solution  winds around the pole of the potential. Note
that for a given topological sector the BPS saturated solution is the
one with lowest spatial action density and thus it is stable
\cite{ShifmanDvali}. In what follows we will restrict ourselves to
$n=1$.

After solving eq.~(\ref{BPS}) we now show that the Maxwell equations
with the source current
%*******
\eqb
\label{cur}
j_\mu\equiv ie\,\delta_{\mu 0}\ \left[{\bar\phi}^{(1)}\stackrel{\rightarrow}{\cal D}_\tau \phi^{(1)}-
\left(\overline{\phi^{(1)}\stackrel{\leftarrow}{\cal D}_\tau}\right) \phi^{(1)}\right]
\eqe
%*******
can be solved by the pure gauge $A_\mu=\delta_{\mu 0}A(\tau)$, in
accord with  local Lorentz invariance of the ground state. Maxwell's
equations in a gravitational background  read \cite{WeinbergB}
%*********
\eqb
\label{ME}
\pd_\mu\left[\,\sqrt{\tilde{g}} F_{\mu\nu}\,\right]=j_\nu\ ,
\eqe
%*********
where $\tilde{g}$ is the determinant of the metric tensor with
Euclidean signature. From eqs.~(\ref{cur},\ref{ME}) it is clear that
$A_\mu$ being pure gauge implies the vanishing of ${\cal D}_\tau
\phi^{(1)}$. Hence all we have to do is to show in a conveniently
chosen gauge that with the background configuration of
eq.\,(\ref{BPSsol}) for $n=1$ there exists a pure gauge of the form
$A_\mu=\delta_{\mu 0}A(\tau)$ with ${\cal D}_\tau \phi^{(1)}=0$. This
is most obvious in unitary gauge $\phi^{(1)}=|\phi^{(1)}|$ where according to
eq.~(\ref{BPSsol}) we have $\pd_\tau\phi^{(1)}=0$. Hence, ${\cal
D}_\tau \phi^{(1)}=0$ for $A(\tau)\equiv 0$.

We now show that the back reaction of the vacuum gauge field onto the
vacuum scalar dynamics is zero, at least in the sense of an
average. Since the $\tau$ dependence of the solution
eq.\,(\ref{BPSsol}) resides only in its phase, observables do not
depend on $\tau$. So averaging over the cycle leaves the potential
invariant. Although the gauge covariant BPS equations 
%*****
\eqb
\label{fse}
{\cal D}_\tau\phi=\bar{V}^{1/2}\,,\ \ \ \ \overline{{\cal D}_\tau\phi}={V}^{1/2} 
\eqe
%*******
are not satisfied by the above field configurations, their cycle
average is. Using eqs.\,(\ref{BPS}) the cycle average can easily be
shown to vanish 
%*********
\eqb
ie \int_0^\beta d\tau\, A_\tau \phi^{(1)}(\tau)=i\frac{2\pi}{\beta}\sqrt{\frac{M^3\beta}{2\pi}}\,
\int_0^\beta d\tau\, \e^{2\pi i \tau/\beta}=0 \; .\
\eqe
%*********
Since we have 
%********
\eqb
\label{masses}
\frac{m_A}{m_\phi}=\frac{e}{\sqrt{6}}\,\left(\frac{|\phi|}{M}\right)^3 
\eqe
%******* 
scalar modes are much heavier than the vector bosons within the entire
relevant range of $\phi$ amplitudes. Hence, the inflaton is a slow
variable as compared to the vector field. Therefore, the employed
procedure to solve the vacuum dynamics of the model is a
Born-Oppenheimer approximation. Also, possible scalar excitations have
mass much larger than $T$ for $|\phi|<|\phi_c|$ which allows for a 
classical treatment of the ground state dynamics \cite{hof5}.

We are now in a position to perform a trivial analytical continuation
of the Euclidean action to Minkowskian signature by letting
$\tau\rightarrow -it$. The kinetic term
$\overline{{\cal D}_\tau\phi}{\cal D}_\tau\phi$ remains
zero. Therefore, we derive a $T$ dependent cosmological constant
%**********
\eqb
\label{La}
\La=\La(T)\equiv V(|\phi^{(1)}|^2)=2\pi M^3 T\ .
\eqe
%*********
This result may be puzzling for the reader. We have derived a finite heat 
capacity of the vacuum without ever referring to a density
matrix\footnote{We thank Daniel Chung for bringing up this objection.}. 
Let us explain this fact. It is useful to think of the vacuum at temperature $T$ 
as a medium with a constant heat capacity which is carried by microscopic degrees of
freedom being of a very different nature than the (irrelevant) quanta of the
inflaton field ($m_\phi/T\sim 15.4$!). If we could probe with a resolution much greater 
than $T$ we would observe the microscopic dynamics directly and 
most probably find effects that even break local Lorentz invariance. However, 
this high resolution is not available and so the 
ground state at resolution $T$ deserves to be called a vacuum. 
The memory of high-scale quantum dynamics is 
contained in the shape of the effective potential \cite{hof5}. A way
to picture this is an analogy with the condensed state represented by
a piece of metal. In this case the macroscopic property heat capacity
is explained by quasidegrees of freedom --- phonons and valence electrons. 
In an effective, macroscopic description these underlying degrees of freedom are 
integrated out.

\subsection{Gravitational sources}

We now investigate the implications of the above vacuum dynamics for
the energy balance in a Friedmann universe at high $T$. Cosmology will
not only be affected by the vacuum structure but also by thermal
excitations. The Higgs mechanism induces a mass $m_A=e|\phi|$ for the
gauge field. If $m_A\ll T$ the spectrum of these excitations can be
approximated by black body radiation, and we can take the mass $m_A$
into account perturbatively when calculating the energy density
$\ep_A$ and the pressure $p_A$. In principle, there is the possibility
of inflaton excitations. Their mass $m_{\phi}=\pd^2_{|\phi|}
V(|\phi|^2)$ is much larger than $T$ for $|\phi|$ away from $|\phi|_c$
\cite{hof5}. Hence, we may disregard scalar excitations (Section 4). 
Considering three polarization states of the massive
vector bosons, expanding up to order $m_A^2$, and expressing mass in
terms of temperature by virtue of eq.\,(\ref{BPSsol}), we obtain
%********
\eqb
\label{epT}
\ep_A=\frac{\pi^2}{10}T^4-\frac{e^2}{16\pi} M^3 T\ ,\ \ \ \ 
p_A=\frac{\pi^2}{30}T^4-\frac{e^2}{24\pi} M^3 T\ . 
\eqe
%********
The matter of the universe is a 
perfect fluid with energy density $\ep$ and pressure $p$ given as
%*******
\eqb
\label{pf}
\ep=\ep_A+\La\ ,\ \ \ \ \ p=p_A-\La\ ,
\eqe
%*********
where the equation of 
state $\La=\ep_\La=-p_\La$ of a cosmological constant has been
used. Using $\Lambda\equiv V(|\phi^{(1)}|^2)$ and eq.~(\ref{epT}), we
can write $\ep$ and $p$ as functions of $\Lambda$, namely,
%*****
\eqb
\label{epL}
\ep=\frac{\Lambda^4}{160\pi^2 M^{12}}+\kappa_1 \Lambda\ ,\ \ \ \ 
p=\frac{\Lambda^4}{480\pi^2 M^{12}}-\kappa_2 \Lambda\ , 
\eqe
%********
where
%********
\eqb
\label{kap}
\kappa_1\equiv 1-\frac{e^2}{32\pi^2}\ ,\ \ \ \kappa_2\equiv 1+\frac{e^2}{48\pi^2}\ .
\eqe
%********
We are now in a position to solve the Friedmann equations. 
Here, we are in particular concerned with the equation governing 
energy-momentum conservation \cite{WeinbergB}
%********
\eqb
\label{FrieI}
\frac{d}{da} \left(\ep\,a^3\right)=-3\,a^2\, p\ .
\eqe
%******** 
Upon use of eq.~(\ref{epL}) together with eq.\,(\ref{FrieI}) we obtain
%********
\eqb
\label{Lp}
\frac{d}{da}\Lambda=-\frac{\Lambda}{a}\,
\frac{\frac{\Lambda^3}{40\pi^2M^{12}}+3(\kappa_1-\kappa_2)}{\frac{\Lambda^3}{40\pi^2M^{12}}+
\kappa_1}\ .
\eqe
%*********

\subsection{Attractor property of inflaton dynamics}

Equation~\ref{Lp} is solved by
%********
\eqb
\label{sol}
\frac{a}{a_0}=\left(\frac{\Lambda}{\Lambda_0}\right)
^{\frac{\kappa_1}{3(\kappa_2-\kappa_1)}}\left(\frac{\Lambda^3+120\pi^2(\kappa_1-\kappa_2)M^{12}}
{\Lambda_0^3+120\pi^2(\kappa_1-\kappa_2)M^{12}}\right)
^{\frac{2\kappa_1-3\kappa_2}{9(\kappa_2-\kappa_1)}}\ ,
\eqe
%********
which cannot be inverted analytically. In Fig.\,2 we show
$\frac{a(\Lambda)}{a_0}$ for $\La_0/M^4=10^2,10^4,10^7$ and
$e=0.1,0.01$. With $M\sim 10^{13}$ GeV (see next section), which
corresponds to a temperature at inflation of about $T\sim
\frac{M}{2\pi}\sim 10^{12}$ GeV, considering the linear dependence of
$\La$ on $T$, and accounting for the fact that for $T\gg M$ the energy
density of the universe is dominated by radiation, $\ep\sim T^4$. The
initial condition $\La_0/M^4=10^7$ reflects Planck scale physics
($M_P\sim 10^{19}$ GeV). 
%***************
\begin{figure}
%\vskip3mm
\vspace{6.2cm}
\includegraphics{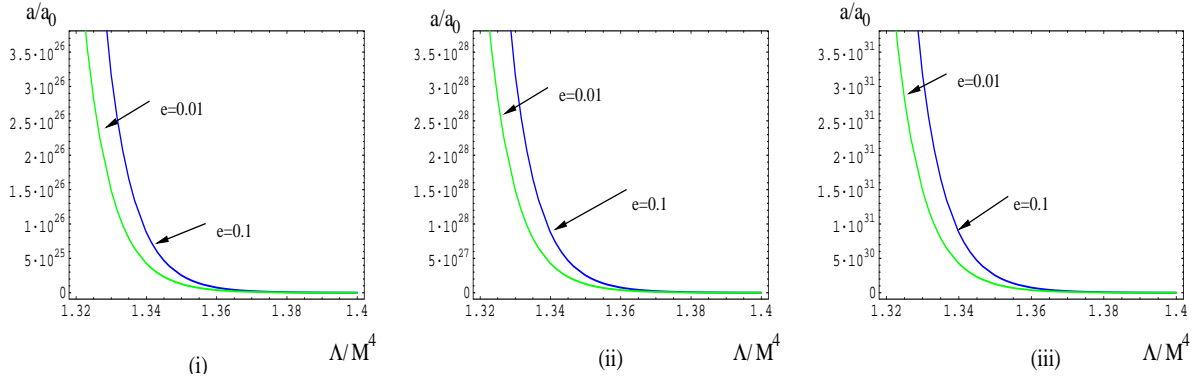}
\caption{The scale factor as a function of the 
cosmological constant. Cases $(i)$, $(ii)$, and $(iii)$ refer to the initial conditions $\La_0/M^4=10^2$, 
$\La_0/M^4=10^4$, and $\La_0/M^4=10^7$, respectively.} 
\end{figure}
%*************** 
According to Fig.\,2 there is no strong dependence on the value of
the gauge coupling $e$. $a/a_0$ has undergone 60 e-foldings, a
phenomenological requirement imposed to sufficiently smooth out the
inhomogeneities existing prior to inflation. The corresponding points
$\La/M^4$ turn out to be between 1.34 and 1.44 for the three very
different initial conditions! So the (inflationary) regime, where the
cosmological constant dominates the energy density, does practically
not depend on the initial conditions prescribed at the borderline of
applicability of our model. We demand that inflation be terminated at
$\La/M^4=1.4$ by a transition to the center symmetric regime. Setting
$V(|\phi|_c^2)=\Lambda=1.4 M^4$, yields $N=34$. We will work with this
value in the following.

\section{Cosmic evolution in thermal equilibrium}
\label{sec:evolution}

\subsection{Flat space solution}

In this section we evaluate the evolution of the scale factor $a$. For
a spatially flat universe ($k=0$) we seek a solution to the Friedmann
equation
%**********
\eqb
\label{FrieII}
H^2(t)\equiv\left(\frac{\dot{a}(t)}{a(t)}\right)^2=\frac{8\pi}{3}\,G\, \ep(a)\ . 
\eqe
%************
Far away from the pole in the RHS of eq.~(\ref{sol}) we may neglect
the constant terms $120\pi^2(\kappa_1-\kappa_2)M^{12}$ and can then
analytically invert eq.~(\ref{sol}), yielding
%******
\eqb
\label{soll}
\La=\La_0\left(\frac{a}{a_0}\right)^{-1}\ .
\eqe
%****** 
Inserting this result into the RHS of the Friedmann equation
(\ref{FrieII}) and using eq.\,(\ref{epL}), we have
%********
\eqb
\label{FrieIIA}
\left(\frac{\dot{a}}{a}\right)^2=\frac{8\pi}{3}\,G\,
\left[\frac{\La_0^4}{160\pi^2M^{12}}\left(\frac{a}{a_0}\right)^{-4}+
\La_0\kappa_1\left(\frac{a}{a_0}\right)^{-1}\right]\ .
\eqe
%********
For high $T$, where radiation dominates the energy density, we may
neglect the term $\propto\left(\frac{a}{a_0}\right)^{-1}$ in
eq.~(\ref{FrieIIA}). This yields the usual decelerated expansion due
to a relativistic gas
%*********
\eqb
\label{relgas}
\frac{a(t)}{a_{0,r}}=\sqrt{2\alpha_r\,(t-t_{0,h})+1}\,,
\eqe
%*********
where 
%**********
\eqb
\label{alh}
\alpha_r\equiv\sqrt{\frac{G}{60\pi}}\frac{\La_0^2}{M^6}\ .
\eqe
%**********

There is an intermediate regime, where $M^4 \gg \La < 10M^4$ or
equivalently, $\frac{M}{2\pi} \ll T < \frac{5M}{\pi}$. The point of
equality of radiation and vacuum energy is at $\frac{\La}{M^4}\sim
11.6$. For $\La<10M^4$ we may neglect the radiation energy still using
the scaling law $\La\propto a^{-1}$ obtained for the radiation
dominated regime. As far as the duration of this regime is concerned,
this approximation  is likely to yield an upper bound because it
underestimates the value of $\La$ close to the inflationary regime
where $\La$ dominates by far the radiation term in the energy
density. The solution to eq.~(\ref{FrieIIA}) then reads 
%********
\eqb
\label{vacstarts}
\frac{a(t)}{a_{0,rv}}=\left(\frac{\alpha_{rv}}{2}(t-t_{0,rv})+1\right)^2\,,
\eqe
%***********
where 
%********
\eqb
\label{alrv}
\alpha_{rv}\equiv\sqrt{\frac{8\pi}{3}G\kappa_1\Lambda_0}\ .
\eqe
%********
So there is accelerated expansion.

According to eq.\,(\ref{sol}) quasi-exponential expansion sets in when 
$\La$ approaches the zero of $\Lambda^3+120\pi^2(\kappa_1-\kappa_2)M^{12}$ which blows up the RHS due to 
the negative exponent $\frac{2\kappa_1-3\kappa_2}{9(\kappa_2-\kappa_1)}$.

\subsection{Decoupling of gravity from inflaton dynamics}
\label{sec:decoupling}

We now show that gravity effectively disappears from the Euclidean 
second-order equations for the inflaton field as a consequence of 
BPS saturation \ref{BPS} in the regime of high $T$. 
According to eqs.\,(\ref{RHS}) this would be the case if the ratio 
%*********
\eqb
\label{R}
R\equiv\left|\frac{3H\,V^{1/2}}{\frac{\pd V}{\pd \phi}}\right| 
\eqe
%********
was much smaller than unity. Neglecting the small vector mass in the radiation dominated era, we have 
%******
\eqb
\label{Hr}
|H_r|\sim\sqrt{\frac{8\pi}{3}G\frac{\pi^2}{10}T^4}\sim
\sqrt{\frac{4}{5}}\pi T^2/M_P\ .
\eqe
%*******
On the other hand, $|V^{1/2}|=\sqrt{2\pi M^3T}$. With eq.~(\ref{Hr}) we then obtain
%*******
\eqb
\label{fric}
|3H\,V^{1/2}|=3\sqrt{\frac{8}{5}\pi^3 M^3T}\,T^2/M_P\ .
\eqe
%*******
Substituting the solution of eq.\,(\ref{BPSsol}) into the derivative of the potential, for 
we have ($n=1$) 
%******
\eqb
\label{deriv}
\left|\frac{\pd V}{\pd \phi}\right|_{\phi^{(1)}}=(2\pi\,M\,T)^{3/2}\ .
\eqe
%****** 
Thus
%*******
\eqb
R_r\sim\frac{3\sqrt{\frac{8}{5}\pi^3 M^3T}\,T^2/M_P}{2(2\pi\,M\,T)^{3/2}}=\sqrt{9/5}\frac{T}{M_P}\ .
\eqe
%******** 
If $T$ is of order $M_P$ this ratio is of order unity. So decoupling of gravity becomes effective 
if the initial temperature is smaller than the Planck mass. Since the dynamics possesses an 
attractor property in the sub-Planckian regime an implicit 
assumption for our model to be viable is that Planckian 
physics drives the universe towards temperatures lower than $M_P$. 

Let us now consider the ratio $R$ for the inflationary era. There, $|H|$ is determined 
by the vacuum energy, and we have
%*******
\eqb
\label{Hi}
|H_i|\sim\sqrt{\frac{8\pi}{3}G 2\pi M^3 T}\sim
\sqrt{16\pi}M^{3/2}T^{1/2}/M_P\ .
\eqe
%*******
So together with eqs.\,(\ref{fric},\ref{deriv}) and $T\sim M/2\pi$ we obtain
%*******
\eqb
R_i\sim 2\sqrt{18}\frac{M}{M_P}<10^{-3}\ 
\eqe
%******** 
for $M$ being smaller than the GUT scale $\sim 10^{16}$ GeV. 
Thus gravity effectively decouples from the inflaton dynamics 
during and well before inflation.

\subsection{A numerical example}

We are now in a position to demonstrate a particular inflationary scenario. 
Let us assume that the Hubble 
parameter $H$ is quasiconstant along a 
time interval $\Delta t_i\sim 10^{-30}\,$s. 
Imposing that inflation should at least 
generate 60 e-foldings,
%********
\eqb
\label{M}
\Delta t_i H=60=10^{-30}\times\frac{3}{2}\times10^{24}\,\mbox{GeV}^{-2}\times 2.4\times 10^{-19}\times M^2\ ,
\eqe
%********
we obtain $M\sim 1.3\times10^{13}\,$GeV. This is considerably lower than the scale of 
grand unification $\sim10^{16}\,$GeV, and hence the density of topological 
defects potentially arising during 
GUT phase transitions does get diluted to immeasurability during inflation. We use this value for $M$ to estimate 
the time scales of the two regimes preceding inflation. 
For definiteness let us assume Planck scale initial 
conditions $\La_0=10^7\,M^4$. With $M=1.3\times10^{13}\,$GeV this yields 
%********
\eqb
\label{alhn}
\alpha_r=\sqrt{\frac{G}{60\pi}}\frac{\La_0^2}{M^6}\sim 1.01\times10^{20}\,\mbox{GeV}\ .
\eqe
%******** 
According to eq.\,(\ref{relgas}) the time interval $\Delta t_r$ 
belonging to the radiation dominated phase is given as
%*********
\eqb
\label{dh}
\Delta t_r=\frac{1}{2\alpha_r}\left[\left(\frac{a_r}{a_{0,r}}\right)^2-1\right]\ ,
\eqe
%********
where $a_r$ and $a_{0,r}$ denote the scale factors at the end and 
the beginning of this regime, respectively. 
From a numerical evaluation of eq.\,(\ref{sol}) we obtain 
$\frac{a_r}{a_{0,r}}\sim 10^6$ for $\La\sim 10\,M^4$ marking 
the end of the epoch of radiation domination. Using this together with 
eqs.\,(\ref{dh},\ref{alhn}), we have 
%*********
\eqb
\label{dhn}
\Delta t_r\sim 3.3\times 10^{-33}\,\mbox{s}\ .
\eqe
%*********
So on a logarithmic scale the radiation dominated epoch 
before inflation lasted rather long.

Let us now define the epoch of spontaneous gauge symmetry breaking being driven by a mixture of 
radiation and vacuum energy. Evolving from $\La=10\,M^4$, where for Planckian initial 
conditions we had ${a_r}/a_{0,r}\sim 10^6$, 
we define the start of inflation by $\La=2\,M^4$ corresponding to $a/a_{0,r}\sim 10^{13}$. 
This yields ${a_{rv}}/a_{0,rv}\sim 10^7$. The expansion 
parameter $\alpha_{rv}$ can be computed as
%*********
\eqb
\label{alrvn}
\alpha_{rv}=\sqrt{\frac{8\pi G}{3}\La_0}\sim 1.3\times 10^8\,\mbox{GeV}\ .
\eqe
%********* 
Solving eq.\,(\ref{vacstarts}) for $\Delta t_{rv}$, we obtain
%*******
\eqb
\label{drv}
\Delta t_{rv}=
\frac{2}{\alpha_{rv}}\,\left(\sqrt{\frac{a_{rv}}{a_{0,rv}}}-1\right)\ .
\eqe
%********
Combining eqs.\,(\ref{alrvn},\ref{drv}) and using 
${a_{rv}}/a_{0,rv}\sim 10^7$, we have
%********
\eqb
\Delta t_{rv}\sim 3\times 10^{-29}\,\mbox{s}\ .
\eqe
%******** 
This is a little more but rather close to the assumed duration of 
inflation of $\Delta t_i\sim 10^{-30}\,$s. 
However, as was argued above, this must be an artifact of the approximation 
made leading to eq.\,(\ref{drv}). 

We now can compare the results obtained in the above, semi-analytic fashion 
with a numerical result. Fig.\,3 shows the computed time evolution of the scale factor. 
It is seen that the radiation approximation worked well. The numerically evolution of $a$ 
in the $rv$ regime, however, is stronger than our analytical estimate.  
%***************
\begin{figure}
%\vskip3mm
\vspace{9.0cm}
\includegraphics{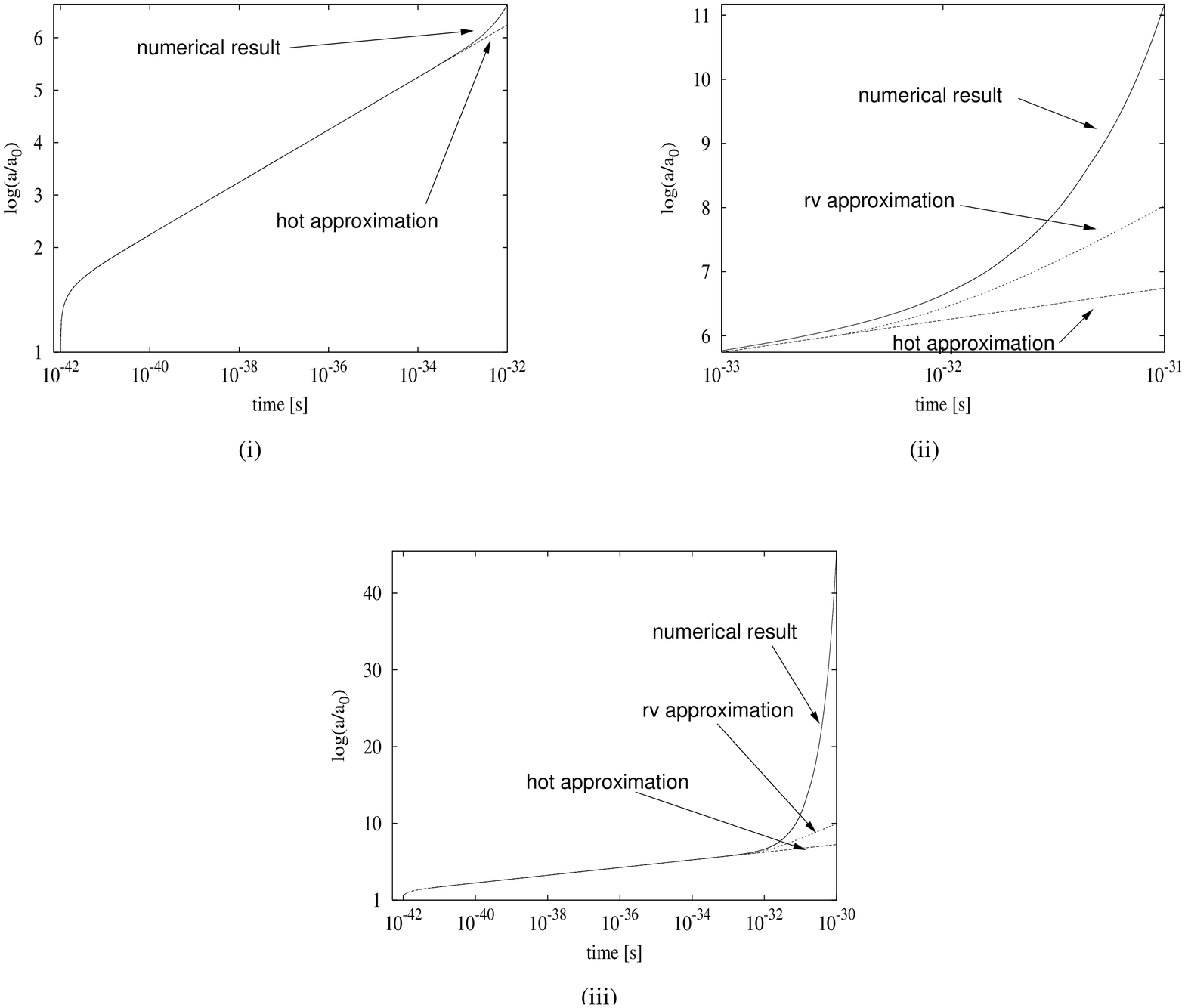}
\caption{Comparison of semi-analytical approximation and numerical 
results for the regimes of radiation ($i$) and of radiation-vacuum mixture ($ii$) 
in a spatially flat universe. The evolution of the scale factor from Planckian initial conditions to 
inflationary cosmology is shown in ($iii$).} 
\end{figure}
%***************

\subsection{Closed universes need not collapse.}

Here we address the fate of a closed universe ($k=+1$ in eq.\,(\ref{Hubble})). 
In the case of a closed or an open universe the square of the 
Hubble parameter $H$ contains an extra term and is given as
%*********
\eqb
\label{Hubble}
H^2=\frac{8\pi G}{3}\left(\frac{\Lambda^4}{160\pi^2 M^{12}}+
\kappa_1 \Lambda-\frac{3k}{8\pi G a^2(t)}\right)\ . 
\eqe
%**********
For a closed universe not to collapse in the course of its evolution the RHS 
of eq.\,(\ref{Hubble}) must remain positive. 
Using radiation scaling $\La=\La_0\frac{a_{0,r}}{a}$, $G^{-1}=M_P^2$, and the 
following relations valid for the Planckian regime 
%*******
\eqb 
\label{Pl} 
\La_0=2\pi T_0 M^3=2\pi M_P M^3\,, \ \ \ \ \ \ M\sim 10^{-6} M_P\ ,
\eqe
%********
we can cast this condition into the form
%**********
\eqb
\label{min}
f(x)\equiv\sqrt{\frac{4}{15} \pi^3 x^2+\frac{16}{3}\pi^2\kappa_1\times 10^{-18} \frac{1}{x}}>(a_{0,r} M_P)^{-1}\ .
\eqe
%**********
Thereby, we have defined $0< x\equiv\frac{a_{0,r}}{a}\le 1$. 
Eq.\,(\ref{min}) yields a lower bound on 
the radius $a_0$ of the universe at Planckian energy density provided that radiation scaling is applicable. 
For estimation purposes we can safely set $\kappa_1=1$. 
The only extremum of the function $f(x)$ in $0\le x\le1$ is a minimum at 
$x_{m}\sim 1.5\times 10^{-6}$. This corresponds to $f(x_m)\sim 7.3\times 10^{-6}$ yielding 
the following lower bound for $a_0$
%*********
\eqb
\label{ab}
a_{0,r}>\frac{1}{7.3}\times 10^6 M_P^{-1}\sim 10^{5} M_P^{-1}\ .
\eqe
%*********
Note that the above scaling assumption for $\La$ is 
justified a posteriori due to the results of the last section stating 
that at $x_{m}\sim 1.5\times 10^{-6}$ 
radiation scaling is an excellent approximation. 
Thus, we have proved that closed universes do not collapse in the 
course of their evolution provided their radius $a_{0,r}$ at 
Planckian density is larger than some critical value. Fig.\,4 shows numerical evolutions 
for open, flat, and closed universes. Thereby, the initial radius $a_{0,r}=1/8.65\times10^{6}M_P^{-1}$ 
has been chosen very close to the critical value. The reader may object to the large hierarchy of 
$\sim 10^5$ between $a^{-1}_{0,h}$ and $M_P$. This hierarchy has its origin in the theory of 
matter and that of gravity being governed by the 
hierarchical mass scales $M$ and $M_P$, respectively, at the 
very regime of separation of the two theories. Only a consistent unified theory of 
gravity and matter can explain this fact. Needless to say, such a theory does not yet exist.  
 
We stress at this point that in a conventional 
scenario, where the matter of the universe close to 
Planckian density is composed of radiation and a $T$ independent $\La$, there always will be a collapse. 
This is due to the second term under the square root in eq.\,(\ref{min}) being a constant. 
The global minimum of the corresponding function $f_c(x)$ then is at $x=0$, which invalidates the derivation of 
the inequality corresponding to eq.~(\ref{min}).
%***************
\begin{figure}
%\vskip3mm
\vspace{5.5cm}
\includegraphics{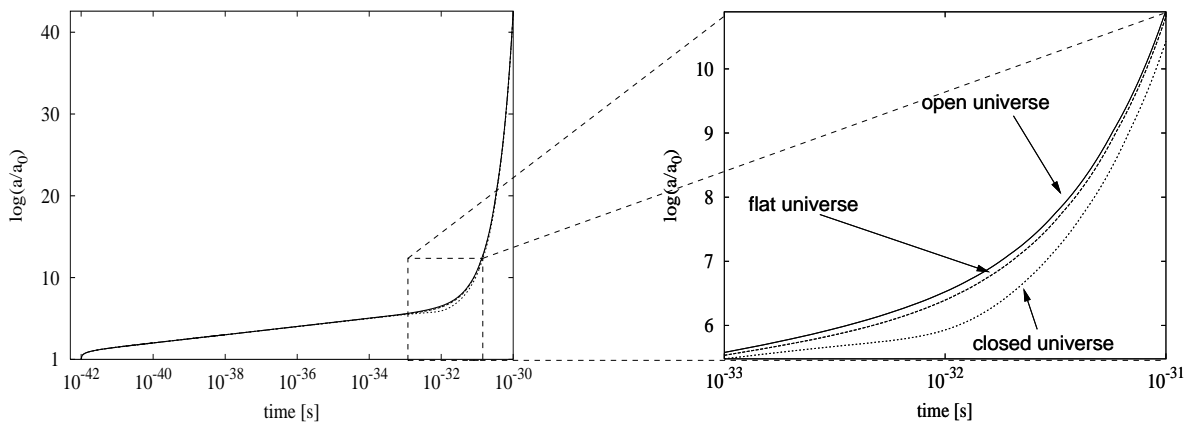}
\caption{Evolution of open, flat and closed universes. We have set 
$a_{0,r}=1/8.65\times10^{6}M_P^{-1}$ which is very near to the 
critical radius for a closed universe.} 
\end{figure}
%***************

\section{The universe after inflation}
\label{sec:postinf}
\subsection{Termination of inflation and reheating}

Quasi-exponential expansion is terminated at 
the point where the inflaton amplitude starts to fluctuate putting an end to 
vacuum dominance. In our model this effect sets in at $|\phi|=|\phi|_c$. 
Even the solutions to the Euclidean 
mean field equations yield a time dependent inflaton amplitude indicating the 
breakdown of thermal equilibrium. Therefore, the dynamics 
can not be treated in Euclidean signature anymore. On the other hand, the 
spontaneously broken gauge symmetry is reduced to a spontaneously broken $Z_{N+1}$ symmetry. 
Thus, gauge bosons cease to get emitted 
from the vacuum. The gauge theory is reduced to a theory of a complex scalar field moving mainly along one of 
the unit root directions and hence to a good approximation can be 
chosen real. To give an estimate on the number of Hubble times, which the subsequent regimes of tachyonic preheating and reheating last, 
we consider spatially homogeneous fluctuations obeying to linear order
%********
\eqb
\label{fluctuations}
\ddot{\delta\phi}+3H\dot{\delta\phi}-m^2(\phi_0)\delta\phi=0\, ,
\eqe
%********
where $m^2\equiv\left.\frac{\pd^2 V}{\pd\phi^2}\right|_{\phi=\phi_0}$. The mean field $\phi_0$ denotes a 
solution to the full equation of motion. Therefore, $\delta\phi \sim e^{mt}$. 
Numerically, $m$ is of the order of several $M$ during 
tachyonic preheating. On the other hand, the Hubble parameter $H$, which can be taken inflation valued, 
is $H\sim\frac{M}{M_P}M\sim 10^{-6} M$.  If we do not allow 
for a fine-tuning of the initial conditions for $\delta\phi$ the typical time interval $\Delta t_{tac}$ the system is in 
the tachyonic regime is about $10^{-6}$ inflationary Hubble times. During 
the subsequent regime of reheating (positive mass squared) $\phi$ performs damped oscillations about its vacuum value $\phi=M$. 
If there were only a few oscillations then this regime would last 
again only about $10^{-6}$ Hubble times since the frequency of oscillation should be comparable to the 
mass of excitations $NM=34\,M$ 
\cite{hof5}. Let us compare this picture with numerical simulations of similar situations 
not assuming spatial homogeneity of $\phi$.
  
On the lattice it was found 
recently \cite{Bellido,KofmanLinde,Parry1} (and references therein) that the 
tachyonic instability of the field $\phi$ staying close to the 
symmetry conserving maximum of its potential at inflation converts most of the 
inflation valued $\La$ into matter due to tachyonic preheating. 
One must distinguish between $\phi$ independent and $\phi$ 
dependent curvature at this maximum. In the former case 
the energy release during tachyonic preheating was 
found to be so efficient that a single 
oscillation of the inflaton field relaxes it to the 
minimum of the potential. In order to arrive at this result, a quantum 
mechanical dispersion of the field amplitude was 
assumed and evolved. 
The occupation numbers of fluctuations at low momenta 
grow exponentially in the early stage of the process. If the potential possesses 
a discrete symmetry then at later stages of the evolution this 
growth spreads to higher momenta due to the formation of domain walls 
and classical wave collisions. It was stressed in refs.\,\cite{KofmanLinde,Parry1} that 
initially large domains grow by ``eating'' 
smaller domains so that the field distribution becomes completely 
asymmetric after some finite time being bounded roughly by 
the mass scales of the potential. This is important since 
the phenomenon of cannibalizing domains excludes domain 
walls within today's horizon being dangerous for present day cosmology. 
Our inflaton potential is qualitatively similar to that of the Coleman-Weinberg model investigated 
in ref.\,\cite{KofmanLinde}. Likewise, the $Z_2$ symmetric potential of this model 
possesses a $\phi$ dependent curvature which vanishes 
at the initial point $\phi_i$ of the process. The following results were obtained 
in ref.\,\cite{KofmanLinde}: Due to vanishing initial curvature 
individual fluctuations are quasistable 
so tachyonic growth does not happen. However, there is instanton mediated 
quantum tunneling towards the regime where perturbative, tachyonic 
fluctuations drive the transition along the lines of what was said above. 
Therefore, domain growth 
by ``eating'' should proceed in analogy to pure tachyonic preheating 
once the transition to the regime of tachyonic 
fluctuations is achieved through nonperturbative effects. 

So qualitatively we obtain the same estimate for the duration of the regimes of 
tachyonic preheating and reheating. 
On the other hand, we may easily estimate how long a co-moving mode, which 
exits the horizon during inflation, 
is outside the horizon after inflation: 
The condition is
%*********
\eqb
\label{cohe}
H=\left.(Ha)\right|_{entry}\frac{1}{a_0}=\left.(Ha)\right|_{entry}\frac{1}{a^i_0}\frac{a^i_0}{a_0}\ ,
\eqe
%*********
where $a_0$ and $a^i_0$ are the scale factors at horizon exit and end of inflation, respectively, 
and $H$ is the Hubble parameter at inflation. We write $\frac{a^i_0}{a_0}=\e^n$. 
For the radiation dominated epoch after reheating 
we obtain 
%*******
\eqb
\label{H}
\left.H\right|_{entry}=\frac{2M^2\sqrt{\frac{8\pi}{3}G}}{2\sqrt{\frac{8\pi}{3}G}M^2(t-t_0^i)+1}\ ,
\eqe
%*******
where $t_0^i$ is the end of inflation. From eqs.\,(\ref{cohe},\ref{H}) we obtain
%********
\eqb
t-t_0^i=\frac{1}{2}(4\e^{2n}-1)H^{-1}\sim H^{-1}\ .
\eqe
%******** 
This is to be contrasted with the time scale $\sim 10^{-6} H^{-1}$ 
tachyonic preheating and reheating takes. Thus it is appropriate to 
say that termination of inflation and reheating proceed instantaneously.

\subsection{Adiabatic density perturbations due to the inflaton field?}

The usual procedure to take into account adiabatic density 
perturbations originating from the inflaton field is the following: 
In the presence of a reservoir quantum fluctuations of the field $\phi$ are 
composed of vacuum fluctuations, 
which are always renormalized away 
\cite{LindeB}, and excitations of momentum $k$. A mode with co-moving momentum  
$k_c=\frac{a}{a_0}k$ exits the horizon during inflation at $\frac{a}{a_0}H=k_c$. If we have $m\ll H$ for the mass $m$ 
of this mode and if there is a 
sufficient amount of inflation after horizon-exit the quantum fluctuation 
is frozen into a classical, Gaussian perturbation with the root of the mean-square given in the 
massless limit as \cite{Lythbook}
%*********
\eqb
\label{phisi}
\delta \phi(k)\equiv\frac{H}{2\pi}\ ,
\eqe
%********
where $H$ is the (quasiconstant) Hubble parameter at inflation. 
If the particular mode enters the horizon well after 
the reheating epoch its corresponding density perturbation 
becomes the seed for a curvature 
perturbation. This, in turn, is responsible for 
structure formation at the associated physical length scale 
\cite{Lythbook}. The density contrast $\frac{\delta \rho(k)}{\rho}$ during inflation 
can be estimated as \cite{LindeB}
%*********
\eqb
\label{dencon}
\frac{\delta \rho(k)}{\rho}=\frac{\delta V(\phi(k))}{V(\phi)}\ .
\eqe
%**********
Due to our thermodynamical approach the situation is rather different here: The 
condition $m\ll H$ for slow-roll of the field $\phi$ in a treatment based on Minkowskian signature 
is not satisfied. On the contrary, we have $m\sim 10^6 H\gg H$. 
Nevertheless, during inflation $|\phi|$ rolls very slowly due to a 
very slowly varying temperature.   

Since the energy density of the radiation is 
suppressed by a factor of about $(2\pi)^{-4}$ as 
compared to the vacuum energy density we do 
expect the density fluctuations in the radiation part to 
be of no significance. Let us therefore focus on scalar fluctuations. 
How relevant are these fluctuations during 
inflation which is characterized by a 
temperature $T_i\sim \frac{M}{2\pi}$? The occupation number is of the Bose type
%**********
\eqb
n^s_B(k)=\frac{1}{\exp\left[\frac{\sqrt{k^2+m_i^2}}{T}\right]-1}\ ,
\eqe
%**********
where $m_i$ denotes the mass of scalar excitations 
during inflation. We have
%*********
\eqb
\label{m,T}
m_i\sim\sqrt{6}\,\frac{M^3}{\phi_c^2}\sim\sqrt{6}\,M\ .
\eqe
%********* 
An upper bound for $n^s_B(k)$ during the part of inflation, 
where $m_i$ can be estimated according to eq.~(\ref{m,T}), is
%*********
\eqb
\label{ephi}
n^s_B(k)\le\frac{1}{\exp\left[\frac{m_i}{T_i}\right]-1}\sim 2\times 10^{-7}\ .
\eqe
%*********
Hence, there are practically no fluctuations of $\phi$. 
However, shortly before inflation terminates (${\pd_{|\phi|}^2V}=0$) the mass of fluctuations reduces 
down to values comparable with $T_i$, and excitations are possible. To estimate 
the magnitude of field perturbations possibly seeding the formation of large-scale structure 
following \cite{LindeB} we write  
%********
\eqb
\label{sflu}
\la (\delta \phi)^2\ra=\frac{1}{(2\pi)^3}\int\frac{d^3k}{\sqrt{k^2+m^2}}n(k)\ ,
\eqe
%**********
where $n(k)$ denotes the occupation number of 
the state with spatial momentum $k$, in our case $n=n_B$. 
Eq.\,(\ref{sflu}) is to be interpreted as a weighted sum over 
all excitable modes that may leave the horizon before inflation ends. 
If we assume that the mass $m$ remains roughly 
constant at value $T_i$ during this 
last stage of inflation an evaluation of $n_B$ at horizon exit $k=H$ 
yields $n_B\sim 1/2$. Moreover, assuming that the universe expanded $e^n$ times 
from the first exit of an excited mode to the end of inflation, we obtain the 
following estimate
%*******
\eqb
\label{est}
\la (\delta \phi)^2\ra=\frac{1}{4\pi^2m}\int_{a^{-1}H}^{H}dkk^2=
\frac{1}{4\pi^2}\left[1-e^{-3n}\right]\frac{H^3}{m}<\frac{1}{4\pi^2}\frac{H^3}{m}\ .
\eqe
%******** 
An upper bound for the corresponding density contrast is
%********** 
\eqb
\label{dens}
\frac{\delta T}{T}\sim\frac{\delta\rho}{\rho}\equiv\left.
\frac{\pd V/\pd\phi}{V}\,\delta{\phi}\right|_{\phi=\phi_c}\sim 10^{-9}\ .
\eqe
%**********
This is much too low to explain the measured anisotropy $\frac{\delta T}{T}\sim 10^{-5}$ 
of the CMBR. Therefore, the required density perturbations do not 
originate from the fluctuations of the 
inflaton field in our model. Recently, it was proposed in \cite{Lyth} 
that the spatial curvature perturbations required for the formation 
of the large scale structure can originate from a light 
scalar field not driving inflation. Our model has to be supplemented with 
such a mechanism for the generation of density perturbations. However, 
it is conveivable that a consensate of composite bosons originating from 
fundamentally charged matter may play the role of this light
scalar field. We hope to address this issue in a forthcoming publication. 

There is a final, rather important observation concerning 
isothermal density fluctuations. These fluctuations are associated with a much smaller 
mass scale than the field driving and terminating inflation. Typical candidates for the generation of 
isothermal fluctuations are the Goldstone modes of the spontaneous 
breakdown of a continuous, global symmetry \cite{Freese}. 
Due to the assumption of pure underlying gauge dynamics 
we only have a spontaneous breakdown of a $Z_{N+1}$ symmetry during tachyonic preheating. 
However, (a) this symmetry originated from a gauge symmetry \cite{Wilczek}, and (b), it is not continuous. Therefore, 
Goldstone modes associated with the termination mechanism of 
inflation do not exist in our minimal model.

\section{Summary and outlook}
\label{sec:concl}

We proposed a model for cosmic inflation based on the thermal
equilibrium dynamics of a higgsed $U(1)$ gauge theory which, in turn,
represents an effective description of underlying $SU(N+1)$ pure gauge
dynamics. This was motivated by a possible description of fundamental,
non-Abelian, and unhiggsed gauge theories in terms of Abelian Higgs
models \cite{dualsc,hof5}. Identifying the Higgs field with the
inflaton, we constructed the corresponding scalar potential such that
the theory admits BPS saturated periodic inflaton solutions. The dynamics following from this potential represents
a concrete realization of Krauss' and Wilczek's proposal that gauge
symmetries masquerade as discrete subgroups when tested with low
energies \cite{Wilczek}. Solving the vacuum gauge field equation in
this background according to a Born-Oppenheimer approximation, implied
a vanishing, gauge invariant kinetic term in the action. Therefore, we
obtained a temperature dependent cosmological constant upon trivial
analytical continuation to Minkowskian signature as the sole
contribution from the scalar sector to the total stress energy of the
universe. The Friedmann equations adapted to this set-up yielded
cosmic inflation at a scale being independent of initial
conditions. The $T$ dependence of $\La$ prevents closed universes from
collapsing provided the initial radius is large enough. We considered
the mechanisms responsible for a termination of inflation and thereby
observed that the point at which the gauge symmetry starts to
masquerade as a $Z_{N+1}$ symmetry coincides with a noticeable
violation of thermal equilibrium. It is stressed at this point that
the estimate for the maximal number of e-foldings during thermal
inflation as it was performed in ref.\,\cite{LythStewart} does not
apply to our model. In \cite{LythStewart} it was assumed that 
temperature decreases as $\sim \exp[-Ht]$. On the contrary, 
temperature remains constant during inflation in our approach. 
This is a direct consequence of the vacuum itself carrying a
nonvanishing heat capacity. Recall that the determination of the scale
$M\sim 10^{13}$\,GeV in section 3.3 originated from demanding that
inflation is to generate about 60 e-foldings at $t\sim 10^{-30}$\,s. 
Towards the end of inflation, excitations become first massless and
later tachyonic. The conventional mechanism for the generation of
density perturbations from quantum fluctuations of the inflaton field
during inflation does not apply to our model. The reason is that the
mass of scalar excitations during the bulk of inflation is much
larger than the Hubble parameter which is in contrast to the usual
slow-roll paradigm. However, one may invoke the recently advocated
curvaton \cite{Lyth} to generate density perturbations independently
from the inflaton dynamics. 

The rapid decay of the old vacuum is accompanied by the generation of
new particles whose dynamics is subject to the prevailing $Z_{N+1}$
symmetry. This symmetry strongly suppresses the decay of the old
vector bosons \cite{Wilczek}, \cite{Cohn}. Hence, these particles of
mass $10^{11}$ -- $10^{13}$ GeV could survive to the present. 
They contribute to the dark matter content and originate ultra-high
energy cosmic rays beyond the Greisen-Zatsepin-Kuzmin bound. This
possibility was first discussed in ref.\,\cite{Rubakov}. Due to their
high mass the relic vector bosons may also function as considerable
dark matter sources. 

We speculate that the cosmologically relevant (equilibrium) dynamics
following the termination of inflation and particle creation is
effectively described by a member of the same class of Abelian Higgs
models labeled by the parameters $(e,M,N)$. The mass scale involved
would be lower and therefore a sequence of milder and milder
inflations is induced. It is tempting to interpret the emission of the
CMBR as the particle creation stage following the last inflation. The
vacuum structure being established would then be associated with the
electromagnetic interaction. With $\Omega_\La\sim 0.7$ one would 
expect that the corresponding inflaton field $\phi_{em}$ was close to
the inflationary regime\footnote{accelerated expansion was measured
\cite{exp}}. Therefore, one could estimate the scale $M_{em}$ by using
the limiting formula $|\phi_{em}|\sim M_{em}$ and
$T\sim\frac{M_{em}}{2\pi}$. However, there is at least one good reason
to reject this possibility: A bound on the photon mass is
$m_\gamma<10^{-16}\,$eV \cite{PDB}. Using the electromagnetic coupling
$\alpha_{em}=\frac{e_{em}^2}{4\pi}=\frac{1}{137}$, one obtains from
$m_\gamma=e_{em}|\phi_{em}|\sim e_{em}M$ that
$M_{em}\lsim3\times10^{-16}\,$eV. Therefore, the energy density
attributed to this vacuum is $\La_{em}\sim M^4\sim
30\times10^{-64}\,(\mbox{eV})^4$. On the other hand, the temperature
of the CMBR is $\sim3\,\mbox{K}=0.25\times10^{-3}\,$eV. This
corresponds to $M_{em}\sim 2\pi T\sim 1.6\times10^{-3}\,$eV which is
in stark contradiction to the value deduced from the photon mass
bound. So if the present accelerated expansion is again driven by a
higgsed, Abelian gauge theory it probably has nothing to do with the
interactions of the standard model of particle theory. This, however,
leaves us with an exciting perspective regarding today's measured
ratio of $\Omega_\La$ and $\Omega_{\rm{dark\; matter}}$. Assuming that
an otherwise undetected Abelian Higgs model drives today's accelerated
expansion, our scenario easily accommodates this ratio.

\section*{Acknowledgements}    

We would like to thank G. Raffelt for valuable comments on the
manuscript. Interesting conversations with A. Dighe, R. Buras,
M. Kachelrie{\ss}, D. Maison, M. Pospelov, G. Raffelt, D. Semikoz, and 
L. Stodolsky are gratefully acknowledged.

\bibliographystyle{prsty}

\end{document}